\documentclass{ws-p8-50x6-00}

\makeatletter
\newcommand\preprintnote{%
        To appear in Proceedings of 9th International Workshop on Deep
        Inelastic Scattering and QCD (DIS 2001), Bologna, Italy, 27 Apr--1
        May 2001 
}
\newcommand\preprint{
   \def\ps@plain{%
       \let\@mkboth\@gobbletwo
       \let\@oddhead\@empty
       \def\@oddfoot{%
           \lower1pc\hbox{\fbox{%
              \parbox{\hsize}{{\it \preprintnote}}%
           }}
       }%
   }
   \def\ps@simple{%
       \let\@mkboth\@gobbletwo
       \let\@oddhead\@empty
       \def\@oddfoot{\hfill {\bf \thepage}\hfill}%
   }
   \AtBeginDocument{
       \pagestyle{simple}
   }
}
\makeatother
\preprint

\providecommand{\MSbar }{\ensuremath{ \overline{\rm MS} }}

\begin{document}

\title{QCD and Diffraction}

\author{John Collins\footnote{On leave from:
        Physics Department,
        Penn State University, %\\
        104 Davey Laboratory,
        University Park PA 16802
        U.S.A.
     }
}

\address{%
   DESY, Notkestra{\ss}e 85, D-22603 Hamburg, Germany, \\
   {\em and}\\
   II Institut f{\"u}r Theoretische Physik, Universit{\"a}t Hamburg, \\{}
   Luruper Chaussee 149, D-22761 Hamburg, Germany
\\ 
E-mail: collins@phys.psu.edu}

\maketitle

\abstracts{%
  I review some of the main results on hard diffraction, in their relation
  to QCD, and indicate some of the strengths and weaknesses of the
  arguments.  I also make some suggestions for future work.
}

%==============================================================
\section{Introduction}

This talk summarized some ideas on how well we know properties of
diffractive and small-$x$ physics as a consequence of QCD.  The primary
results concern the separation of phenomena on different scales and in
different rapidity ranges, and it this separation or factorization that
enables predictions and useful models to be made despite our ignorance
about non-perturbative QCD:
\begin{enumerate}
\item Ordinary hard-scattering factorization for $F_2^{\rm(inclusive)}$,
  jet production, etc.
\item The same for diffractive hard processes.
\item BFKL, etc.
\item $k_T$-factorization, CCFM, etc.
\item Dipole model, etc.
\end{enumerate}
There is also the subject of exclusive diffraction and skewed pdf's, which
I will not dicuss here. I will also not discuss the issue of Regge theory
and Regge factorization, except indirectly in the context of the BFKL
equation.  Regge factorization does not follow in any demonstrated way
from QCD, if one includes non-perturbative phenomena, and in the context
of hard scattering, it clearly fails in its simplest form.  However, this
is an important area of physics that needs to be properly elucidated.

In the general area of diffraction, a cultural divide exists between those
who work in the Breit frame (in DIS) and those who work in the target rest
frame.  The first group use a parton density language \cite{diff_fact} and
the second tend to use the dipole model \cite{dipole}.  The two frames lead
to very different intuitions.  But QCD is the same for all of us, and
progress is to be made by understanding the compatibility of the two
views.

My remarks are confined to lepto-production processes, even though the
same issues extend to many other processes.

%==============================================================
\section{Parton viewpoint: hard-scattering factorization}

\begin{figure}
  \centering
  \includegraphics[scale=0.4]{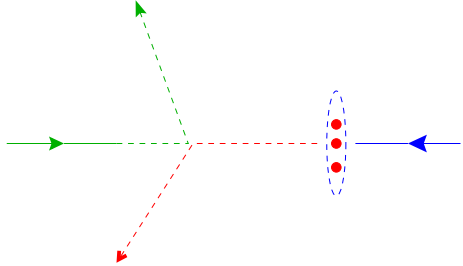}
  \caption{DIS: electron from left, proton from right.}
  \label{fig:DIS-pic}
\end{figure}

The main ideas of the parton method are best understood in coordinate
space in the Breit frame --- Fig.\ \ref{fig:DIS-pic}.  A highly
time-dilated and Lorentz contracted proton arrives from the right.  For
numerical illustration, suppose that $Q^2 = 10 \, {\rm GeV}^2$ and
$x=2\times10^{-4}$.  The hard interaction occurs with one constituent over
a scale $1/Q$, about $0.1\,{\rm fm}$.  The struck quark has an energy of
$1.6\,{\rm GeV}$, and its scale of non-perturbative interactions is
$5\,{\rm fm}$, due to time-dilation.  The valence interactions of the
proton occur on a much longer scale $10^4\,{\rm fm}$.

This separation of scales leads to factorization, for $F_2$ and many other
cross sections.  The leading power result is
\begin{equation}
  \label{eq:factn}
  \mbox{hard-scattering coefficient} \otimes \mbox{parton density}.
\end{equation}
This also applies \cite{diff_fact} to diffractive processes, where the
parton density is replaced by a diffractive parton density (or extended
fracture function), which is a number density of partons conditional on a
diffracted proton in the final state.

The intuitive picture is clear, and is fully and provably realized in QCD.
Results include: (a) the DGLAP evolution of pdf's, (b) the need for
higher-order corrections (non-leading logarithms) to the hard scattering
and DGLAP kernels, and (c) methods to obtain these corrections from
Feynman graphs.  The successes of these methods is well-known.

However, there is a myth that has arisen and that is, I believe, hindering
further progress.  This is that hard scattering involves scattering of
on-shell partons.  Clearly this is false, since the partons are confined.
Discussions predicated on the myth are misleading, if not
incomprehensible, for students.  New results in the field depend on a more
exact treatment of parton kinematics, which can only be done correctly if
the myth is removed.

Simple factorization breaks down when $Q$ is too small, and ``higher
twist'' corrections come into play.  More interestingly, it breaks down
when $1/x$ is sufficiently large that ``saturation'' effects matter.  Even
if saturation does not yet occur, there are enhancements in higher order
corrections at small $x$.

%==============================================================
\section{Small $x$ logarithms and saturation}

At small $x$ two problems arise.  One is that in gluon ladders there are
integrals over the rapidities of intermediate gluons that lead in
coefficient functions and in the DGLAP kernels to logarithms of $x$.
Low-order perturbation theory for these quantities is no longer accurate.
Possible remedies are (a) resummation of the large logarithms using the
BFKL equation, (b) use of sufficiently many higher orders of perturbation
theory, and (c) use of $k_T$ factorization.  These are in principle
compatible, but rely on approximations, and the $k_T$ factorization
approach is probably the most unifying method.

A more fundamental breakdown of the methods occurs at sufficiently small
$x$ that saturation occurs.  The intuitive picture leading to
factorization requires that the different partons in the target hadron be
transversely separated at the moment of the hard scattering.  This fails
when the partons overlap transversely, when the packing fraction
\begin{equation}
\label{eq:saturation}
  \rho = \frac{\mbox{Area of partons}}{\mbox{Area of proton}}
       = \frac{\mbox{number of parton} /Q^2}{1/R^2},
\end{equation}
becomes larger than unity.  Simple parton ideas fail, and a breakdown of
the DGLAP prediction of the $Q$ fails even if calculations are carried to
high orders of perturbation theory (non-leading logarithms).  

In applying this idea quantitatively, one should know that at small $x$, 
\MSbar{} pdf's do not provide a good measure of the number of partons,
since they are not equal to an appropriate integral of an unintegrated
parton density. 

%==============================================================
\section{$k_T$ factorization}

A fundamental way \cite{kt_fact} of tackling these and other problems is
to use ``unintegrated parton densities'' $P(x,k_T)$ that give the $k_T$
dependence of the partons.

At small $x$ and high $Q$ the LLA suggests that ladder-like graphical
contributions dominate.  As one goes from the virtual photon to the
proton, the parton $x$ increases and the parton $k_T$ generally
decreases.  By locating where the first large gap in one of these
variables occurs one obtains a factorization:
\begin{equation}
  \label{eq:kt_fact}
  d\sigma = \mbox{coefficient function} \otimes P(x,k_T) .
\end{equation}
There are no large ranges of kinematics in the coefficient, so it is
perturbatively calculable.  But if the kinematic gap below it is in $x$,
we must explicitly take account of parton $k_T$; hence the use of the
unintegrated pdf.  

A proper treatment needs care.  For example, non-ladder graphs are
important, but the need for most of them can be removed by the use of
light-cone gauge.  However, the light-cone gauge introduces divergences
when the rapidities of virtual gluons go to $-\infty$, a region that is
completely unrelated to the physics.  There is no cancellation between
real and virtual gluons in the situation we are considering, so it is
compulsory to provide a cut-off: $P(x,k_T,y_c)$.  Effectively, the BFKL
and CCFM equations provide an equation for the dependence on the cut-off
$y_c$.

Unfortunately, the cut-off is not explicit in public derivations, with the
obvious resulting problems.  

This impacts on the discussions of saturation.  It is essential to discuss
this in terms of $P(x, k_T, y_c)$.  The relation to the \MSbar{} pdf's
needs to be explicitly quantified.  Many years ago, in other processes,
Soper and I \cite{CS} gave a suitable concrete definition, with the
rapidity cutoff explicit.  The definition of quark-parton state includes a
certain amount of ``wee glue'', and the equation for the $y_c$ dependence
is a kind of renormalization-group equation with respect to the amount of
wee glue.

Also, it is not obvious to me how the published derivations go beyond the
leading-logarithm approximation; in particular, at sufficiently
non-leading logarithmic accuracy, there is more than single-gluon exchange
across kinematic gaps in $x$, contrary to what was needed to obtain Eq.\ 
(\ref{eq:kt_fact}).  This matters if one wishes to use $k_T$-factorization
as a property of full QCD, including non-perturbative effects.

%==============================================================
\section{Dipole model}

A totally different is the dipole model \cite{dipole}.  Its intuition
comes from the target rest frame, where the time-scale of the virtual
photon is much greater than the time for it to cross the target.  If $Q^2
= 10 \, {\rm GeV}^2$ and $x=2\times10^{-4}$, as before, the photon energy
is $2.5\times10^4\,{\rm GeV}$ and the $q\bar q$ component of the photon
has a formation distance of around 500\,fm, whereas the proton is just
1\,fm across.

Clearly, it is a sensible first approximation to treat the photon's $q\bar
q$ component as frozen in transverse size while it crosses the target.
This results in a lot of useful phenomenology.  In particular inclusive
$F_2$ and diffraction are related by applying the optical theorem to the
dipole cross section.

It is important that the basic dipole, with a size of order 1\,fm, is in
the ``aligned jet'' configuration.  One particle of the dipole has much
higher energy than the other.  The high-energy particle is the struck
quark, which gives the high $p_T$ jet in the HERA laboratory frame.  The
other, ``slow'' quark is actually in the parton density: it is the quark
of fractional momentum $x$ before it is struck, and the order of events
for this quark is opposite than in the Breit frame.  This change of time
ordering between frames implies that the quark propagates over a {\em
  space}-like distance, and therefore that intuitive ideas may be somewhat
misleading.

The correspondence with the factorization ideas is as follows.  If the
transverse momentum in the pair is small, then the dominant situation (to
power-law accuracy) is the asymmetric aligned-jet configuration, with a
size of $1/k_T$; this corresponds to the LO parton model.  Provided we use
the leading logarithm approximation in $x$, time dilation of the dipole
system ensures that scattering occurs over a short distance scale, and the
dipole idea is really valid.  

If $k_T$ is large ($\sim Q$) we have a small size configuration, with the
parton momenta not being greatly different.  The lifetime of the state
remains long (500\,fm in the example).  However, scattering can occur (in
non-leading logarithm in $x$) by emission of a slow gluon.  We then have a
3-body system, which is effectively an octet dipole, consisting of a gluon
well separated from a small octet $q\bar q$ system.

Intermediate configurations are involved in giving the correct DGLAP
evolution. 

Clearly ``QCD corrections'' give non-trivial modifications to the dipole
model, and it is questionable how well it is preserved beyond the
perturbative domain for anything but qualitative purposes.  The dipole
model is most at home in a leading logarithm approximation.  Quantitative
comparisons of data and dipole model predictions can be quite tricky
particularly to determine whether saturation occurs.

%==============================================================
\section{Relation of dipole-model to factorization}

Clearly, there must be a relation between dipole model and pdf formalism.
However, it is a non-trivial relation, because the time sequence of events
is reversed between the natural reference frames for the two methods, and
because of the need to treat non-leading logarithmic and non-perturbative
effects. Even so, the basic Feynman-diagrammatic view is the same, and the
theory is the same. It is most natural, I think, to try to relate the
dipole cross section to (unintegrated) pdf's.

Some hints as to the details can be found in the work \cite{dipole} of
Frankfurt, Strikman, et al.  One definitely has to ask what happens beyond
the leading approximation.  The factorization point-of-view includes, in
principle, all non-leading logarithms, and its pdf's are fully
non-perturbative.

%==============================================================
\section{Conclusions}

I would like to know the answers to the following questions, which in my
personal opinion have not been completely satisfactorily answered in the
literature: 
\begin{enumerate}
\item What {\em precisely} is the unintegrated pdf $P(x,k_T,y_c)$?
\item Explicitly, what is the rapidity cut-off on virtual gluons?
\item What is the relation to $\sigma_{\rm dipole}(x,r_T)$?
    (What variables? Why?)
\item What is the quantitative relation to the \MSbar{} pdf's?
\item We need a more complete derivation of the space-time structure.
  E.g.: 
  \begin{itemize}
  \item[] Where is diffracted proton formed?
  \item[] What about Regge theory?
  \end{itemize}
\end{enumerate}

%==============================================================

\end{document}